\documentclass[fleqn,usenatbib]{mnras}

\usepackage{newtxtext,newtxmath}

\usepackage[T1]{fontenc}

\DeclareRobustCommand{\VAN}[3]{#2}
\let\VANthebibliography\thebibliography
\def\thebibliography{\DeclareRobustCommand{\VAN}[3]{##3}\VANthebibliography}

\usepackage{graphicx}	
\usepackage{amsmath}	

\usepackage{amssymb}	

\title[HCS Shift at the end of solar cycle 24]{On the Northward Shift of the Heliospheric Current Sheet at the End of Solar Cycle 24}

\author[H. C. Li and X. S. Feng]{
Huichao Li,$^{1,2}$\thanks{E-mail: lihuichao@hit.edu.cn}
Xueshang Feng,$^{1,2}$\thanks{E-mail: fengx@spaceweather.ac.cn}
\\
$^{1}$Shenzhen Key Laboratory of Numerical Prediction for Space Storm, Institute of Space Science and Applied Technology, Harbin Institute of Technology, \\ Shenzhen, China\\
$^{2}$SIGMA Weather Group, Key Laboratory of Solar Activity and Space Weather, National Space Science Center, Chinese Academy of Sciences, Beijing, China
}

\pubyear{2024}

\begin{document}
\label{firstpage}
\pagerange{\pageref{firstpage}--\pageref{lastpage}}
\maketitle

\begin{abstract}
 Since solar cycle 16, the { heliospheric} current sheet (HCS) has been found to be shifted southward during the late declining to minimum phase. However, this trend is broken at the end of solar cycle 24. In this paper, we analyze the shift of the HCS by using information obtained from coronal  model and insitu data provide by the near-Earth OMNI database and the Parker Solar Probe (PSP). Coronal potential field source surface (PFSS) modeling results show that the northward shift is established at the beginning of 2018 and remains stable for about two years. Interplanetary magnetic field data obtained from and within 1 au also support the northward shift, as the southern polarity T appears more frequently than the northern polarity A between 2018-2020. Both model results and insitu observation obtained by PSP imply that the HCS shift is established in the corona, and then propagates into the heliosphere. The quadrupole term still has a significant influence on the formation of the HCS shift.
\end{abstract}



\section{Introduction}
\label{sec:into}
The heliospheric current sheet (HCS) is the boundary enclosing the Sun which separates interplanetary magnetic field (IMF) of opposite { polarities}. As the outward extension of the solar magnetic equator, the HCS is the key structure that sculpts the heliosphere \citep{Smith2001}. Many solar wind parameters, including speed, density and temperature, are organized with respect to the HCS \citep{Fletcher2015}, and the propagation of cosmic ray is significantly influenced by the HCS \citep{Potgieter2013}. 

  The HCS has been found to be persistently shifted southward during the late declining to minimum phase of the solar cycle \citep{Mursula2003}. They analyzed hourly IMF data dating back to 1964, revealing that this southward shift is a persistent structure that is observable in all available solar cycles. They also coined the nickname "the bashful ballerina" for this phenomenon \citep{Mursula2007}.   Further examination of IMF polarity inferred from geomagnetic records has extended the analysis period to pre-satellite times (1926–1955), suggesting that the southward shift has begun as early as solar cycle (SC) 16 \citep{Hiltula2006,Mursula2011}. Supporting evidence for the southward shift comes from cosmic ray detection of Ulysses spacecraft during its first fast latitude scan between 1994 and 1995 \citep{Simpson1996}. Analysis of the Ulysses IMF data also confirms a southward shift of about  $2^\circ$ around solar minima 22/23 and 23/24 \citep{Erdos2010,Virtanen2010}. The southward shift of the HCS is considered to be linked to the north-south asymmetry in the global distribution of the solar wind speed during minimum 22/23 \citep{Tokumaru2015}.

The continuous southward shift of the HCS is considered to be formed in the corona, and then propagates into the heliosphere. One evidence comes from the position of the current sheet (neutral line) of modeled coronal fields between 1976 and 2001, which are extrapolated by the potential field source surface (PFSS) model. Two long southward HCS displacement intervals are found between 1983 and 1986 and between 1992 and 1995, in the late declining phase close to solar minimum \citep{Zhao2005}. Another evidence is revealed by the longitudinally averaged latitude of the white-light streamer structures, which is considered to be formed around the current sheet. Strong tendency for the HCS to be shifted southward by a few degrees is found during 2007--2011, around minimum 23/24 \citep{Robbrecht2012}.

In our previous study \citep{Li2021APJL}, we find that the HCS is shifted northward during 2019--2020.  This ends the nearly centennial southward-shift tendency of the HCS in the late declining to minimum phase of the solar cycle, reversing it to a northward shift, as predicted by \cite{Mursula2011}. While \citet{Li2021APJL} focus on the general character of corona and solar wind around solar minimum, in this study we provide a comprehensive analysis of this northward shift at the end of SC 24, utilizing information provided by model results and insitu measurements.

\section{ model results}
\label{sec:model}

\begin{figure*}
	\centering
	\centerline{\includegraphics[width=0.7\textwidth]{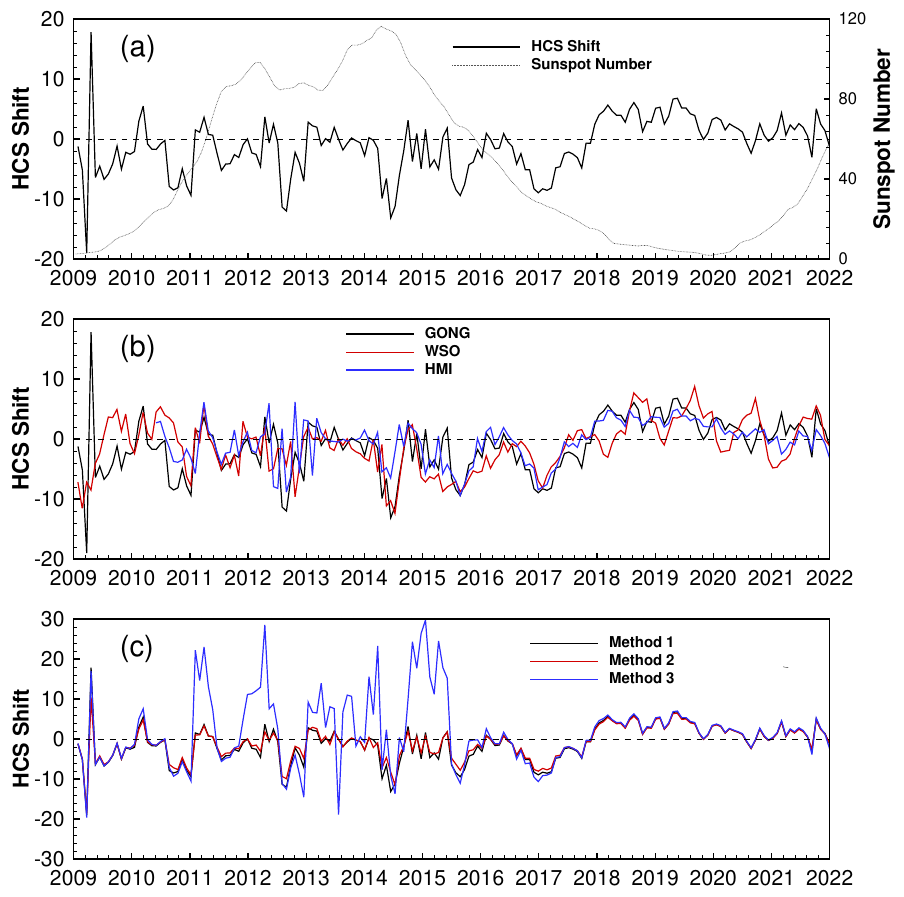}}
	\caption{Shift  {$\lambda$} of the heliospheric current sheet (HCS)  calculated from PFSS model results. Panel (a) shows results obtained with GONG maps, imposed with the sunspot number. Panel (b) compares HCS shift obtained with GONG, WSO and HMI synoptic maps. Panel (C) compares HCS shift calculated with different methods.   }
	\label{fig:PFSS_result}
\end{figure*}

We first use the PFSS model to analyze the HCS shift in the corona. Extensive validation indicates that the PFSS model can reproduce the large-scale coronal structure with reasonable accuracy (see \citet{Mackay2012} and references therein). The PFSS code used in this work is introduced and validated by \citet{Li2021PFSS}. Figure \ref{fig:PFSS_result}(a) presents the shift  {$\lambda$} of HCS from the heliographic equator derived from PFSS model. The { photospheric}  boundary conditions are specified by GONG's integral synoptic maps with their zero-point corrected.  The shift  {$\lambda$} of the HCS is calculated using the algorithm proposed by \citet{Zhao2005} 
\begin{eqnarray}
	\label{equ:shift_zhao2005}
	& \lambda = \lambda_m|cos\delta|  \nonumber \\
	& \lambda_m = \sin^{-1}\left(1-\frac{\Omega_{SS}^{N}}{2\pi} \right)
\end{eqnarray}
In these equations, $\Omega_{SS}^{N}$ is the solid angle of northern magnetic hemisphere at the source surface, which has the same polarity as the dominant polarity in the north polar region of the solar surface. The effective shift of the HCS from the solar dipole equator, $\lambda_m$, is calculated from $\Omega_{SS}^{N}$. The tilt angle of the solar magnetic dipole is calculated by
\begin{equation}
	\label{equ:tilt_zhao2005}
	\delta=\tan^{-1}\frac{\sqrt{g_{11}^2+h_{11}^2}}{g_{10}}
\end{equation}
where $g_{10}$, $g_{11}$, and $h_{11}$ are the spherical harmonic coefficients of the PFSS model  (e.g., Equs (1)-(3) in \citet{Li2021PFSS} ).


As introduced in Section \ref{sec:into}, the HCS shift before SC 24 has been studied in detail. Therefore, we choose to calculate the HCS shift from 2009 to 2021, covering the whole SC 24 and one year after. From Figure\ref{fig:PFSS_result}(a), we can see that before 2018, the HCS shift is dominated by negative values, indicating that the HCS still tends to be shifted southward. However, the extent of shift is quite variable. A southward shift of more than $2^\circ$ can sustain for about half a year at most. Then, as SC 24 enters its declining phase, a stable northward shift is established at the beginning of 2018, and remains stable until the end of 2019, when the solar minimum is reached. As the new solar cycle begins in 2020, the northward shift starts to diminish but is still observable.

The input synoptic maps have a fundamental influence on the PFSS solution \citep[e.g.,][]{Virtanen2019,Nikolic2019,Li2021PFSS}. To further verify the conclusion obtained with GONG maps, we calculate the HCS shift using synoptic maps generated by Wilcox Solar Observatory (WSO) and Helioseismic and Magnetic Imager (HMI) onboard the Solar Dynamics Observatory (SDO) spacecraft (The HMI maps are available since mid-2010). In Figure \ref{fig:PFSS_result}(b), we can see that the overall trends of the HCS shift are in agreement among the three results during the whole SC 24. Between 2018--2020,  the northward shift of the HCS is seen in all of the three results. There seems to be an annual variation in the WSO results, in which the HCS reaches its southernmost/northernmost position in spring/autumn. Such an annual variation has been found in previous studies employing WSO maps \citep[e.g.,][]{Virtanen2016, Li2021APJL}, and is considered as an artificial effect resulting from the { annual variation of the solar tilt (solar-b) angle, which is also known as the Earth's vantage-point effect}. In Table \ref{tab:HCS_shift_PFSS}, we list the average HCS shift during each year and the 3-year interval. The GONG and HMI results indicate that the HCS is shifted northward by about $3^\circ$--$4^\circ$ during 2018 and 2019. In 2020, as the HCS flats back to the heliographic equator \citep{Li2021APJL}, the northward shift reduces to about $1^\circ$. The average northward shift during the 3-year interval is between $2^\circ$-$3^\circ$. Despite the artificial annual variation, the 3-year average northward shift calculated from WSO results does not deviate too much from GONG and HMI. 

\begin{table}
	\centering
	\caption{Yearly and 3-Year HCS shift derived from PFSS results}
	\label{tab:HCS_shift_PFSS}
	\begin{tabular}{lcccr} 
		\hline
		Map & 2018 & 2019 & 2020 & 3-year \\
		\hline
		GONG & 4.0 & 3.9 & 1.2 & 3.0 \\
        HMI  & 3.5 & 3.2 & 1.1 & 2.6 \\
        WSO  & 2.0 & 3.7 & 1.1 & 2.3 \\
		\hline
	\end{tabular}
\end{table}

Besides using Equs (\ref{equ:shift_zhao2005}) and (\ref{equ:tilt_zhao2005}) (Method 1), there are alternative ways to estimate the HCS shift. While still employing Equ (\ref{equ:shift_zhao2005}), \citet{Virtanen2016} use the average of the maximum northern and southern extension of the HCS to calculate the tilt angle of the solar magnetic dipole $\delta$ (Method 2). The longitudinally averaged latitude of the HCS is also employed to represent the shift of HCS (Method 3), when the position of HCS is identified from coronal observation \citep{Robbrecht2012} or magnetohydrodynamics (MHD) modeling result \citep{Li2018}. These two alternative methods are useful for observations and models which cannot provide spherical harmonic coefficients. In Figure \ref{fig:PFSS_result}(c), we compare the HCS shift calculated using the three different methods. While Method 1 and 2 give almost the same estimation, Method 3 results in much larger HCS shift between 2011--2016. During this period, the Sun is in the most active phase of SC 24. The HCS usually has a complex structure and extends to high latitude. In other years, when the solar activity is lower, Method 3 also gives an identical estimation. The comparison indicates that solar activity level should be taken into account when comparing HCS shift calculated by different methods. 



\section{insitu analysis}
\label{sec:insitu_analysis}

\begin{figure*}
	\centering
	\centerline{\includegraphics[width=0.6\textwidth]{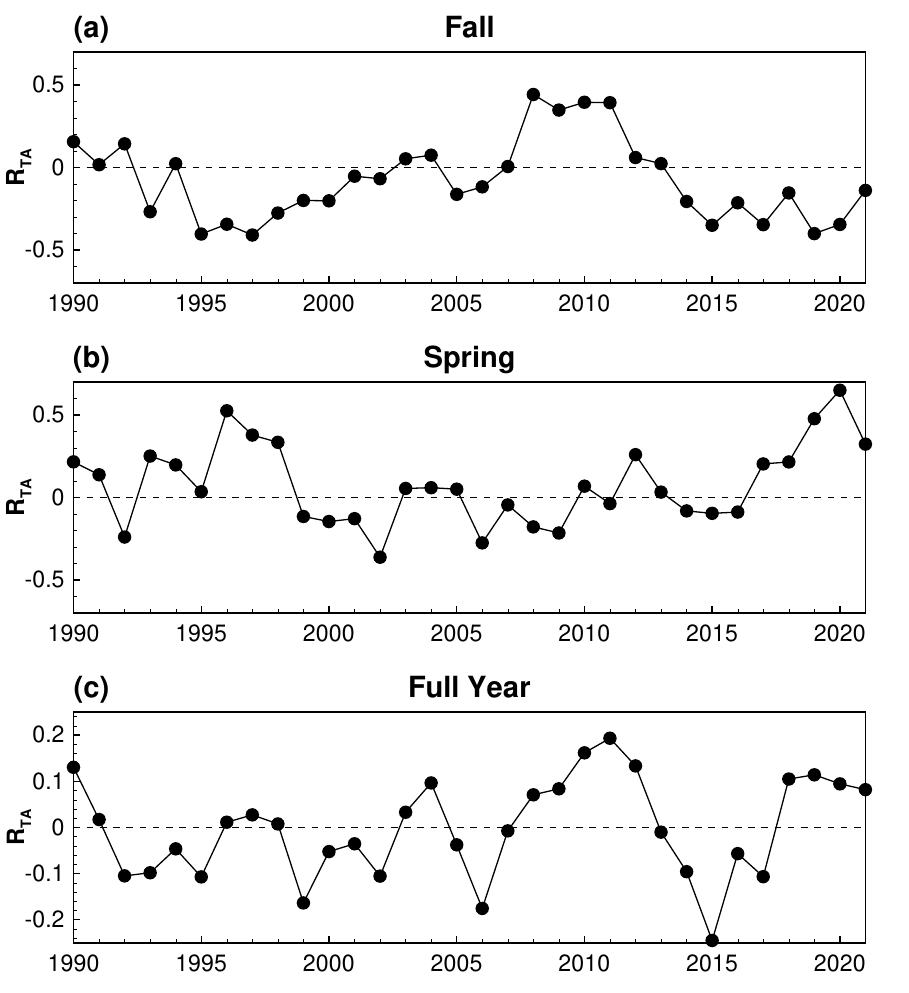}}
	\caption{Seasonal/annual $R_{TA}$ ($R_{TA} = (T-A)/(T+A)$) ratios in 1990--2021 from
		OMNI data. Panel (a): fall months
		(Aug--Oct, north); Panel (b): Spring months (Feb--Apr, south); Panel (c): full years. The horizontal dashed line marks the position where $R_{TA}=0$. }
	\label{fig:insitu_overall}
\end{figure*}

In this section, the shift of HCS is analyzed using information from insitu IMF data. We first focus on the near-Earth data provided by the OMNI database. { Following \citet{Mursula2003},}  the polarity of the IMF data is defined by the plane division, where the toward (T) and away (A) polarity are defined by $B_x > B_y$ and $B_x < B_y$ respectively in the geocentric solar ecliptic (GSE) coordinate system. The relative occurrence difference $R_{TA} = (T-A)/(T+A)$ is calculated to show which polarity dominates.

Figure \ref{fig:insitu_overall} presents the $R_{TA}$ ratio for each Fall (Aug--Oct), Spring (Feb--Apr) and full year.  To compare SC 24 with historical data, the beginning of the analysis interval is expanded to the middle of SC 22. It has been observed that the polarity associated with the Sun's north/south pole dominates the near-Earth insitu data during Fall/Spring around solar minimum. This phenomenon is called the Rosenberg-Coleman (RC) rule \citep{Rosenberg1969} and can be explained by the fact that the Earth obtains its highest northern/southern heliographic latitude during Fall/Spring. In Figure \ref{fig:insitu_overall}, in years around 2019, i.e., the minimum between SC 24/25, the RC rule is valid in both seasons.  The Spring $R_{TA}$ ratio achieves its highest value of 0.650 in 2020, which means that during 82.5\% of the time in that Spring, the Earth encounters T (southern) polarity.  Meanwhile,  an overall dominance of the southern  (T) polarity can be seen from the full year $R_{TA}$ ratios, which have positive values between 2018-2021. All these facts indicate that the HCS is shifted northward at 1 au. However, the shift cannot exceed the $7.2^\circ$ tilt of the solar equator with respect to the ecliptic, as the dominance of the northern (A) polarity is still observable during Fall { \citep{Mursula2003}   } .

\begin{figure*}
	\centering
	\centerline{\includegraphics[width=0.6\textwidth]{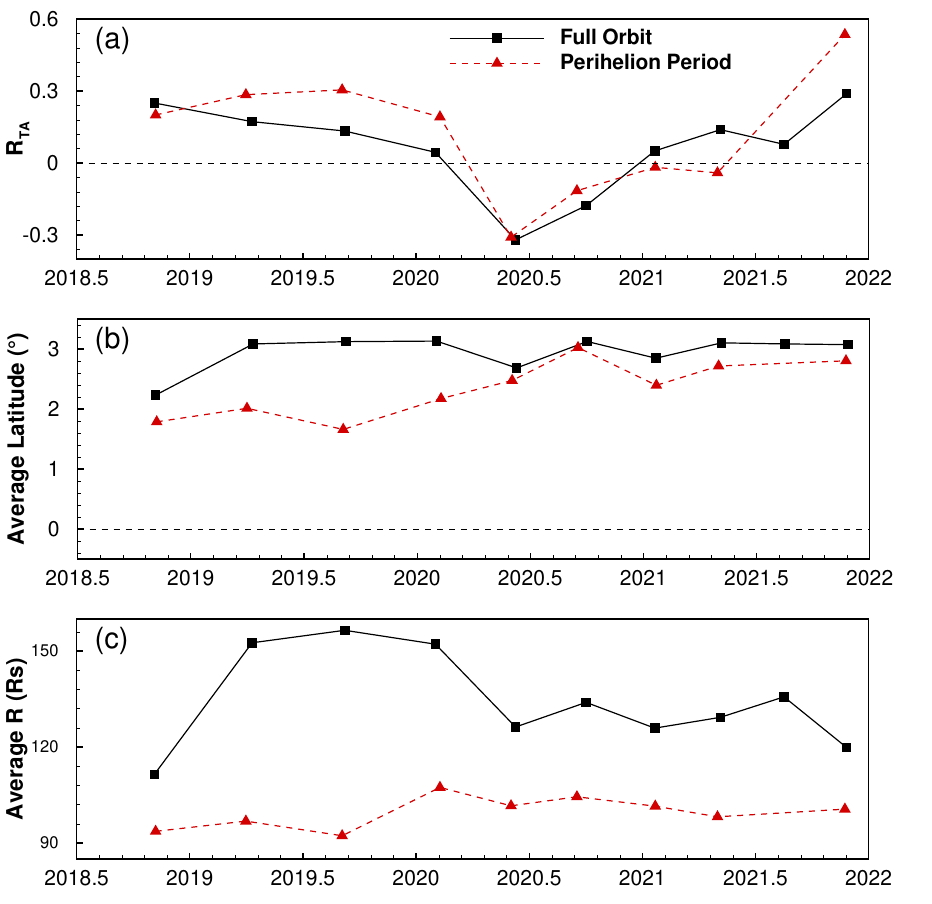}}
	\caption{Panel(a): $R_{TA}$ ($R_{TA} = (T-A)/(T+A)$) calculated from Parker Solar Probe (PSP) data.  $R_{TA}$ of each orbit and perihelion are represented by black and red color respectively. Panel (b) and (c): Average latitude and heliocentric distance (R) of each orbit and perihelion.}
	\label{fig:RTA_PSP}
\end{figure*} 

The Parker Solar Probe (PSP) launched in Aug 2018 provides additional information for the formation and evolution of the HCS shift inside 1 au. In the following, we will look into the $R_{TA}$ ratio calculated by PSP data. The PSP is operated in a highly-elliptical heliocentric orbit with an inclination of about $4^\circ$. From its launch to the end of 2021, the PSP has finished 10 orbits, and its perihelion moves from 35.6 Rs (0.17 au) to 13.3 Rs (0.062 au).  The heliocentric distance, heliolatitude and heliolongitude of the PSP are plotted in Figure \ref{fig:PSP_orbit} in the Appendix.  { Following \citet{MacNeice2009}, the polarity of PSP IMF data is defined by its field angle $\varphi$ in the RT plane of RTN coordinates, where $0^\circ$ is along the radial outward direction}. If  $-135^\circ \leq \varphi \leq 45^\circ$, the IMF polarity is set to be A, otherwise T.

The $R_{TA}$ ratios of each orbit and perihelion are calculated for PSP using the hourly averaged IMF data provided by the Coordinated Data Analysis Web (CDAWeb) service. These ratios are presented in Figure \ref{fig:RTA_PSP}. The IMF data collected between two successive apogees are used to calculate $R_{TA}$ of each orbit. The span of each perihelion period is defined in the following way: the first time that PSP cruises into 0.5 au in each orbit is defined as the beginning, and each perihelion period covers the whole $360^\circ$ longitude in the Carrington coordinate. To give equal weight to each longitude range in the calculation of $R_{TA}$, we distribute the hourly data to longitudinal bins of $1^\circ$, and compute the average IMF value for each bin. The polarity of each bin is determined by the averaged IMF value, and the $R_{TA}$ ratio is then calculated by counting the polarity of each bin. The average heliocentric distance and latitude of PSP during each orbit and perihelion are also presented in Figure \ref{fig:RTA_PSP}.

The PSP cruises in the north of the solar equator during most of its orbit, and only travels to the south hemisphere around its perihelion (See Figure \ref{fig:PSP_orbit} in the appendix). As a result, in Figure \ref{fig:RTA_PSP} the average latitude of each orbit and perihelion period are all positive. If the HCS has no shift, the PSP would encounter more northern polarity (A), and the  $R_{TA}$ would have negative values. However, during 2018 and 2019, the $R_{TA}$ has positive values, indicating the southern polarity (T) dominates and the HCS is apparently shifted northward. Negative $R_{TA}$ ratios are obtained in the last two orbits of 2020, indicating that the northern polarity A appears more frequently. However, due to the north-dominated orbit, we cannot infer southward shift of HCS for this period. As can be seen in the model results (Figure \ref{fig:PFSS_result}), the amount of northward shift apparently decreases during 2020, which may explain the negative $R_{TA}$ in PSP data.  The perihelion data obtains larger $R_{TA}$ values than the full-orbit data, probably due to the lower average latitude. The overall trend of $R_{TA}$ is similar between full-orbit and perihelion data, implying that the HCS shift is established in the corona first, and the evolution in the heliosphere does not  significantly influence the HCS shift.

\begin{figure}
	\centering
	\centerline{\includegraphics[width=\columnwidth]{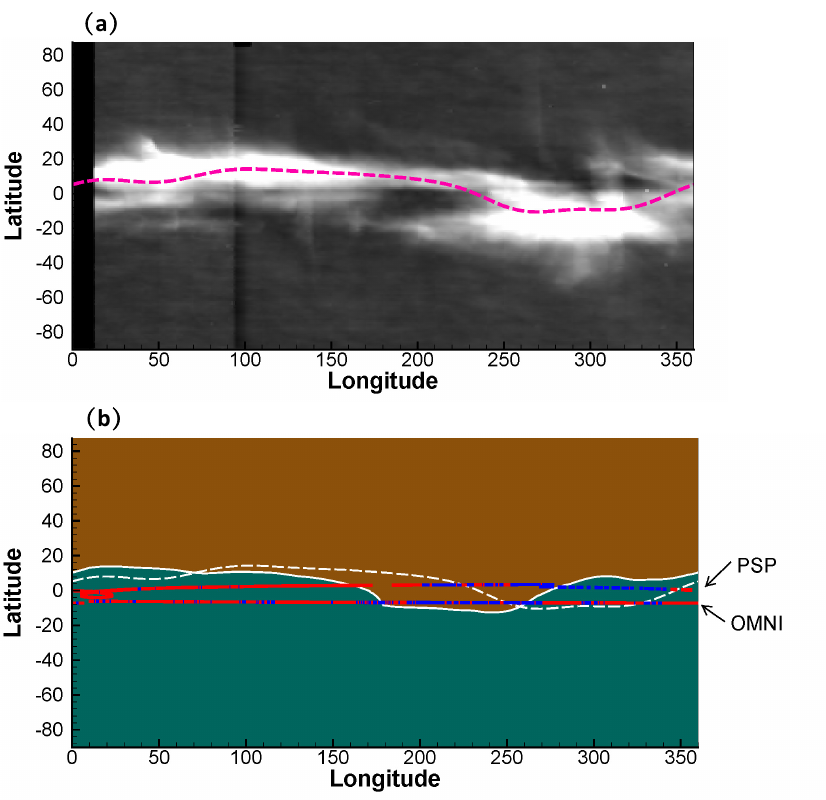}}
	\caption{Comprehensive view of HCS structure in the corona and heliosphere at Carrington Rotation (CR) 2215 { (Mar 12 2019 to Apr 8 2019)}. Panel (a): Synoptic map of Polarized brightness (pB) observation made by SOHO/LASCO C2 at 2.5 $R_s$
		for . The HCS at the source surface (2.5 $R_s$) obtained by PFSS model is marked by {magenta} dashed line. Panel (b): Synoptic map of IMF polarity at 1 au obtained by the heliospheric MHD model. Towards (T) and Away (A) polarity is represented by brown and green color respectively. The solid and dashed white line marks the position of the HCS at 1 au and source surface respectively. The trajectory of Earth (OMNI) and PSP is marked by dots, where T and A polarity is represented by { blue and red } color respectively.     }
	\label{fig:CR2215}
\end{figure}


In Figure \ref{fig:CR2215}, we provide a comprehensive view of HCS structure in the corona and heliosphere, using information obtained from model  and observation. In Figure \ref{fig:CR2215} (a), we present a synoptic map of polarized Brightness (pB) observation made at the height of 2.5 Rs. The HCS derived by the PFSS model at the source surface is superimposed on the pB map by a { magenta} dashed line. { In Figure \ref{fig:CR2215} (b), the synoptic map displays the IMF polarity at 1 AU, which was calculated using a three-dimensional heliospheric MHD model. The computational domain of the MHD model extends from 0.1 au to 1 au. The boundary condition of the MHD model is provided by the coronal PFSS model and other empirical relationships (see  \citet{Li2020} for detail).  In this representation, the T and A polarities are indicated by brown and green colors, respectively. } The position of HCS at the source surface (Rss = 2.5 Rs, dashed white line )  and at 1 au (solid white line) are deduced from PFSS and MHD model results respectively. The trajectory of Earth (OMNI) and PSP around PSP's 1st Perihelion of 2019 is marked by dots, whose color represents the IMF polarity detected ({ blue for T and red for A }). The PSP data obtained at different heliocentric distances are mapped back to the source surface ballistically following the Parker spiral.  The model results are obtained using synoptic maps of CR 2215, whose temporal span {(Mar 12 2019 to Apr 8 2019)} is covered by the PSP data. 

As can be seen from the figure, the position of PFSS-derived HCS agrees reasonably with observed pB bright structures.  {It should be noted that there are cases when the pB bright structures and the HCS are oppositely displaced \citep[e.g.,][]{Crooker1997,Mursula2002}. The possible cause of this separation is the occurrence  of the pseudostreamer, which separates magnetic flux converging from coronal holes of the same polarity and contains no current \citep{Wang2007}. The PFSS model is generally reliable in reproducing the HCS structure. It has been employed to distinguish helmet streamer (with current) and pseudostreamer (without current) \citep[e.g.,][]{Owens2014,Crooker2014,Wang2019}. Therefore, it is reasonable here to assume that pB bright structure enveloping the PFSS-derived HCS in Figure \ref{fig:CR2215} (a) reflects the actual position of the HCS.} Meanwhile, the position of the modeled HCS at the source surface and 1 au  generally match the sector boundary measured by the PSP and OMNI respectively. These comparisons demonstrate the validity of model results. The HCS has a relatively flat shape.  Two factors contributing to the northward shift of the HCS can be identified from these plots. First, the maximum northern extension of the HCS is about $15^\circ$, which is slightly larger than the maximum southern extension of about $10^\circ$. Second, the longitudinal range of the northward shift is larger than that of the southward shift by about $180^\circ$. At 1 au, the southward-shifted HCS is only found between longitude $170^\circ$--$260^\circ$, while other longitudinal range is covered by northward-shifted HCS. The shape and latitudinal extension of the HCS do not change significantly from the source surface to 1 au. A longitudinal shift between the source surface and 1 au can be seen, and such shift is a natural consequence of solar corotation. The closest approach of the PSP to the Sun is achieved between longitude $5^\circ$-$25^\circ$, where the trajectory of the PSP forms a small circle. During this period, the IMF polarity detected by PSP is nearly all positive, even without any significant fluctuation, implying that the HCS is well northward of the PSP, agreeing with the model results.

\section{summary and discussion}
\label{sec:discussion}

In this paper, we find that the HCS is shifted northward during the late declining phase of SC 24. Coronal PFSS modeling results show that the northward shift is established at the beginning of 2018 and remains stable for about two years. During 2020, the northward shift started to diminish. Results obtained from different synoptic map sources agree with each other in general, although the degree of shift has small discrepancies. Insitu IMF polarity data obtained from and within 1 au also support the northward shift, as the southern polarity T appears more frequently than the northern polarity A between 2018-2020. Both model results and insitu observation obtained by PSP imply that the overall trend of HCS shift is established in the corona, and then propagates into the heliosphere.

It is pointed out by \citet{Wang2011} that the dominance of one polarity at the heliographic equator or Earth does not necessarily imply a systematic north-south shift of the HCS. The fluctuation in the nonaxisymmetric photospheric field may also influence the equatorial sector width, which is reflected by the $R_{TA}$ ratio. Nevertheless, as we can see from the model result in Figure \ref{fig:CR2215}, the HCS is confined in the low latitude within $-10^\circ \sim 15^\circ$ region, and has a flat shape during the selected CR. In our previous study \citep{Li2021APJL}, we also find that the HCS remains flat shape around minimum 24/25. If the nonaxisymmetric photospheric field is significant, the flat shape of the HCS would be interrupted by apparent warps.  Therefore, for the period studied in this paper, the nonaxisymmetric photospheric field does not play an important role, and the  $R_{TA}$ ratio calculated from near-equator data is still a valid indicator for HCS shift. 

\begin{figure*}
	\centering
	\centerline{\includegraphics[width=0.75\textwidth]{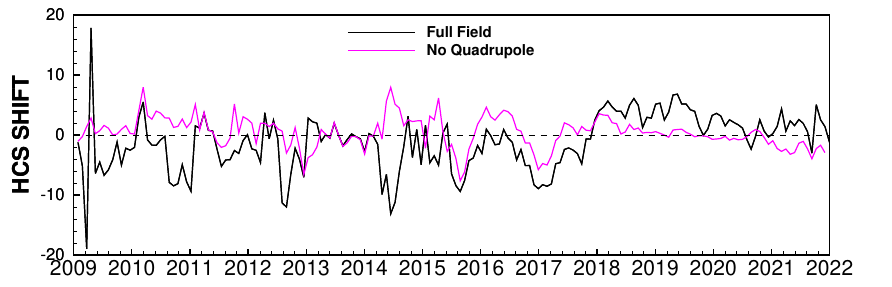}}
	\caption{HCS shift  according to full field (black) and to an expansion where the quadrupole term ($g^2_0$) is neglected (purple).}
	\label{fig:Quadrupole}
\end{figure*}


It has been  inferred that the axial quadrupole term with sign opposite to the dominant dipole plays a crucial role in the formation of HCS shift around solar minimum \citep{Wang2011,Virtanen2014}. This quadrupole term is usually expressed by the $g^2_0$ term in the spherical harmonics form of the PFSS solution.  In Figure \ref{fig:Quadrupole}, we plot the HCS shift calculated according to full expansion (black, same as Figure \ref{fig:PFSS_result} (a)) and to an expansion where $g^2_0$ term is neglected. The northward shift of the HCS during 2018-2020 nearly disappears in the $g^2_0$ neglected results, indicating that the quadrupole term still influences the HCS shift significantly during the late declining and minimum phase of SC 24, when the direction of HCS shift is different from previous cycles.

 As mentioned in Section \ref{sec:into}, IMF information regarding HCS shift is only available since the 1930s. On the other hand, the hemispheric asymmetry in solar wind speed has been studied, utilizing geomagnetic activity data since the 1840s. It is found that the north-south asymmetry in the solar wind has changed its orientation between the asymmetric intervals of the mid-19th century and the late 20th century,  implying that the asymmetry of heliosphere may oscillate at the period of about 200-300 years \citep{Mursula2001}, most likely at the 205-210-year deVries/Suess cycle \citep{Mursula2007}. Considering the close relationship between the latitudinal structure of solar wind speed and the position of the HCS,   \citet{Mursula2011} proposed that, based on the 200-300 year cycle, the orientation of HCS asymmetry would change after the end of solar cycle 23. The northward shift of the HCS observed in this study validates that prediction. Furthermore, it is intriguing to explore the implications of this change on phenomena closely linked to HCS, such as the variation of the cosmic ray, and the distribution of the solar wind plasma properties in the heliosphere.

\section*{Acknowledgements}

The work is jointly supported by the National Natural Science Foundation of China (42030204, 42204174), Shenzhen Natural Science Fund (the Stable Support Plan Program GXWD20220817152453003),  Shenzhen Key Laboratory Launching Project (No. ZDSYS20210702140800001), and Guangdong Basic and Applied Basic Research Foundation (2023B1515040021). The present work is also partially supported by the National Key Scientific and Technological Infrastructure project ``Earth System Science Numerical Simulator Facility'' (EarthLab). This work utilizes GONG data from NSO, which is operated by AURA under a cooperative agreement with NSF and with additional financial support from NOAA, NASA, and USAF. We acknowledge use of NASA/GSFC's Space Physics Data Facility's OMNIWeb and CDAWeb service for providing the observed {\it in-situ} data used in this paper. 

\section*{Data Availability}

 The data underlying this article will be shared on reasonable request to Huichao Li (lihuichao@hit.edu.cn) or Xueshang Feng(fengx@spaceweather.ac.cn).



\bibliographystyle{mnras}
\bibliography{hcs_shift} 

\begin{thebibliography}{}
\makeatletter
\relax
\def\mn@urlcharsother{\let\do\@makeother \do\$\do\&\do\#\do\^\do\_\do\%\do\~}
\def\mn@doi{\begingroup\mn@urlcharsother \@ifnextchar [ {\mn@doi@}
  {\mn@doi@[]}}
\def\mn@doi@[#1]#2{\def\@tempa{#1}\ifx\@tempa\@empty \href
  {http://dx.doi.org/#2} {doi:#2}\else \href {http://dx.doi.org/#2} {#1}\fi
  \endgroup}
\def\mn@eprint#1#2{\mn@eprint@#1:#2::\@nil}
\def\mn@eprint@arXiv#1{\href {http://arxiv.org/abs/#1} {{\tt arXiv:#1}}}
\def\mn@eprint@dblp#1{\href {http://dblp.uni-trier.de/rec/bibtex/#1.xml}
  {dblp:#1}}
\def\mn@eprint@#1:#2:#3:#4\@nil{\def\@tempa {#1}\def\@tempb {#2}\def\@tempc
  {#3}\ifx \@tempc \@empty \let \@tempc \@tempb \let \@tempb \@tempa \fi \ifx
  \@tempb \@empty \def\@tempb {arXiv}\fi \@ifundefined
  {mn@eprint@\@tempb}{\@tempb:\@tempc}{\expandafter \expandafter \csname
  mn@eprint@\@tempb\endcsname \expandafter{\@tempc}}}

\bibitem[\protect\citeauthoryear{{Crooker}, {Lazarus}, {Phillips}, {Steinberg},
  {Szabo}, {Lepping}  \& {Smith}}{{Crooker} et~al.}{1997}]{Crooker1997}
{Crooker} N.~U.,  {Lazarus} A.~J.,  {Phillips} J.~L.,  {Steinberg} J.~T.,
  {Szabo} A.,  {Lepping} R.~P.,   {Smith} E.~J.,  1997, \mn@doi [\jgr]
  {10.1029/96JA03681}, \href
  {https://ui.adsabs.harvard.edu/abs/1997JGR...102.4673C} {102, 4673}

\bibitem[\protect\citeauthoryear{{Crooker}, {McPherron}  \& {Owens}}{{Crooker}
  et~al.}{2014}]{Crooker2014}
{Crooker} N.~U.,  {McPherron} R.~L.,   {Owens} M.~J.,  2014, \mn@doi [Journal
  of Geophysical Research (Space Physics)] {10.1002/2014JA020079}, \href
  {https://ui.adsabs.harvard.edu/abs/2014JGRA..119.4157C} {119, 4157}

\bibitem[\protect\citeauthoryear{{Erd{\H{o}}S} \& {Balogh}}{{Erd{\H{o}}S} \&
  {Balogh}}{2010}]{Erdos2010}
{Erd{\H{o}}S} G.,  {Balogh} A.,  2010, \mn@doi [Journal of Geophysical Research
  (Space Physics)] {10.1029/2009JA014620}, \href
  {https://ui.adsabs.harvard.edu/abs/2010JGRA..115.1105E} {115, A01105}

\bibitem[\protect\citeauthoryear{{Fletcher}, {Cargill}, {Antiochos}  \&
  {Gudiksen}}{{Fletcher} et~al.}{2015}]{Fletcher2015}
{Fletcher} L.,  {Cargill} P.~J.,  {Antiochos} S.~K.,   {Gudiksen} B.~V.,  2015,
  \mn@doi [\ssr] {10.1007/s11214-014-0111-1}, \href
  {https://ui.adsabs.harvard.edu/abs/2015SSRv..188..211F} {188, 211}

\bibitem[\protect\citeauthoryear{{Hiltula} \& {Mursula}}{{Hiltula} \&
  {Mursula}}{2006}]{Hiltula2006}
{Hiltula} T.,  {Mursula} K.,  2006, \mn@doi [\grl] {10.1029/2005GL025198},
  \href {https://ui.adsabs.harvard.edu/abs/2006GeoRL..33.3105H} {33, L03105}

\bibitem[\protect\citeauthoryear{{Li} \& {Feng}}{{Li} \& {Feng}}{2018}]{Li2018}
{Li} H.,  {Feng} X.,  2018, \mn@doi [Journal of Geophysical Research (Space
  Physics)] {10.1029/2017JA025125}, \href
  {https://ui.adsabs.harvard.edu/abs/2018JGRA..123.4488L} {123, 4488}

\bibitem[\protect\citeauthoryear{{Li}, {Feng}, {Zuo}  \& {Wei}}{{Li}
  et~al.}{2020}]{Li2020}
{Li} H.,  {Feng} X.,  {Zuo} P.,   {Wei} F.,  2020, \mn@doi [\apj]
  {10.3847/1538-4357/aba61f}, \href
  {https://ui.adsabs.harvard.edu/abs/2020ApJ...900...76L} {900, 76}

\bibitem[\protect\citeauthoryear{{Li}, {Feng}  \& {Wei}}{{Li}
  et~al.}{2021a}]{Li2021PFSS}
{Li} H.,  {Feng} X.,   {Wei} F.,  2021a, \mn@doi [Journal of Geophysical
  Research (Space Physics)] {10.1029/2020JA028870}, \href
  {https://ui.adsabs.harvard.edu/abs/2021JGRA..12628870L} {126, e28870}

\bibitem[\protect\citeauthoryear{{Li}, {Feng}  \& {Wei}}{{Li}
  et~al.}{2021b}]{Li2021APJL}
{Li} H.,  {Feng} X.,   {Wei} F.,  2021b, \mn@doi [\apjl]
  {10.3847/2041-8213/ac13a6}, \href
  {https://ui.adsabs.harvard.edu/abs/2021ApJ...917L..26L} {917, L26}

\bibitem[\protect\citeauthoryear{{MacNeice}}{{MacNeice}}{2009}]{MacNeice2009}
{MacNeice} P.,  2009, \mn@doi [Space Weather] {10.1029/2009SW000463}, \href
  {https://ui.adsabs.harvard.edu/abs/2009SpWea...7.6004M} {7, S06004}

\bibitem[\protect\citeauthoryear{{Mackay} \& {Yeates}}{{Mackay} \&
  {Yeates}}{2012}]{Mackay2012}
{Mackay} D.~H.,  {Yeates} A.~R.,  2012, \mn@doi [Living Reviews in Solar
  Physics] {10.12942/lrsp-2012-6}, \href
  {https://ui.adsabs.harvard.edu/abs/2012LRSP....9....6M} {9, 6}

\bibitem[\protect\citeauthoryear{{Mursula}}{{Mursula}}{2007}]{Mursula2007}
{Mursula} K.,  2007, \mn@doi [Advances in Space Research]
  {10.1016/j.asr.2007.05.087}, \href
  {https://ui.adsabs.harvard.edu/abs/2007AdSpR..40.1034M} {40, 1034}

\bibitem[\protect\citeauthoryear{{Mursula} \& {Hiltula}}{{Mursula} \&
  {Hiltula}}{2003}]{Mursula2003}
{Mursula} K.,  {Hiltula} T.,  2003, \mn@doi [\grl] {10.1029/2003GL018201},
  \href {https://ui.adsabs.harvard.edu/abs/2003GeoRL..30.2135M} {30, 2135}

\bibitem[\protect\citeauthoryear{{Mursula} \& {Virtanen}}{{Mursula} \&
  {Virtanen}}{2011}]{Mursula2011}
{Mursula} K.,  {Virtanen} I.,  2011, \mn@doi [\aap]
  {10.1051/0004-6361/200913975}, \href
  {https://ui.adsabs.harvard.edu/abs/2011A&A...525L..12M} {525, L12}

\bibitem[\protect\citeauthoryear{{Mursula} \& {Zieger}}{{Mursula} \&
  {Zieger}}{2001}]{Mursula2001}
{Mursula} K.,  {Zieger} B.,  2001, \mn@doi [\grl] {10.1029/2000GL011880}, \href
  {https://ui.adsabs.harvard.edu/abs/2001GeoRL..28...95M} {28, 95}

\bibitem[\protect\citeauthoryear{{Mursula}, {Hiltula}  \& {Zieger}}{{Mursula}
  et~al.}{2002}]{Mursula2002}
{Mursula} K.,  {Hiltula} T.,   {Zieger} B.,  2002, \mn@doi [\grl]
  {10.1029/2002GL015318}, \href
  {https://ui.adsabs.harvard.edu/abs/2002GeoRL..29.1738M} {29, 1738}

\bibitem[\protect\citeauthoryear{{Nikoli{\'c}}}{{Nikoli{\'c}}}{2019}]{Nikolic2019}
{Nikoli{\'c}} L.,  2019, \mn@doi [Space Weather] {10.1029/2019SW002205}, \href
  {https://ui.adsabs.harvard.edu/abs/2019SpWea..17.1293N} {17, 1293}

\bibitem[\protect\citeauthoryear{{Owens}, {Crooker}  \& {Lockwood}}{{Owens}
  et~al.}{2014}]{Owens2014}
{Owens} M.~J.,  {Crooker} N.~U.,   {Lockwood} M.,  2014, \mn@doi [Journal of
  Geophysical Research (Space Physics)] {10.1002/2013JA019412}, \href
  {https://ui.adsabs.harvard.edu/abs/2014JGRA..119...36O} {119, 36}

\bibitem[\protect\citeauthoryear{{Potgieter}}{{Potgieter}}{2013}]{Potgieter2013}
{Potgieter} M.~S.,  2013, \mn@doi [\ssr] {10.1007/s11214-011-9750-7}, \href
  {https://ui.adsabs.harvard.edu/abs/2013SSRv..176..165P} {176, 165}

\bibitem[\protect\citeauthoryear{{Robbrecht} \& {Wang}}{{Robbrecht} \&
  {Wang}}{2012}]{Robbrecht2012}
{Robbrecht} E.,  {Wang} Y.~M.,  2012, \mn@doi [\apj]
  {10.1088/0004-637X/755/2/135}, \href
  {https://ui.adsabs.harvard.edu/abs/2012ApJ...755..135R} {755, 135}

\bibitem[\protect\citeauthoryear{{Rosenberg} \& {Coleman}}{{Rosenberg} \&
  {Coleman}}{1969}]{Rosenberg1969}
{Rosenberg} R.~L.,  {Coleman} Paul~J. J.,  1969, \mn@doi [\jgr]
  {10.1029/JA074i024p05611}, \href
  {https://ui.adsabs.harvard.edu/abs/1969JGR....74.5611R} {74, 5611}

\bibitem[\protect\citeauthoryear{{Simpson}, {Zhang}  \& {Bame}}{{Simpson}
  et~al.}{1996}]{Simpson1996}
{Simpson} J.~A.,  {Zhang} M.,   {Bame} S.,  1996, \mn@doi [\apjl]
  {10.1086/310127}, \href
  {https://ui.adsabs.harvard.edu/abs/1996ApJ...465L..69S} {465, L69}

\bibitem[\protect\citeauthoryear{{Smith}}{{Smith}}{2001}]{Smith2001}
{Smith} E.~J.,  2001, \mn@doi [\jgr] {10.1029/2000JA000120}, \href
  {https://ui.adsabs.harvard.edu/abs/2001JGR...10615819S} {106, 15819}

\bibitem[\protect\citeauthoryear{{Tokumaru}, {Fujiki}  \& {Iju}}{{Tokumaru}
  et~al.}{2015}]{Tokumaru2015}
{Tokumaru} M.,  {Fujiki} K.,   {Iju} T.,  2015, \mn@doi [Journal of Geophysical
  Research (Space Physics)] {10.1002/2014JA020765}, \href
  {https://ui.adsabs.harvard.edu/abs/2015JGRA..120.3283T} {120, 3283}

\bibitem[\protect\citeauthoryear{{Virtanen} \& {Mursula}}{{Virtanen} \&
  {Mursula}}{2010}]{Virtanen2010}
{Virtanen} I.~I.,  {Mursula} K.,  2010, \mn@doi [Journal of Geophysical
  Research (Space Physics)] {10.1029/2010JA015275}, \href
  {https://ui.adsabs.harvard.edu/abs/2010JGRA..115.9110V} {115, A09110}

\bibitem[\protect\citeauthoryear{{Virtanen} \& {Mursula}}{{Virtanen} \&
  {Mursula}}{2014}]{Virtanen2014}
{Virtanen} I.~I.,  {Mursula} K.,  2014, \mn@doi [\apj]
  {10.1088/0004-637X/781/2/99}, \href
  {https://ui.adsabs.harvard.edu/abs/2014ApJ...781...99V} {781, 99}

\bibitem[\protect\citeauthoryear{{Virtanen} \& {Mursula}}{{Virtanen} \&
  {Mursula}}{2016}]{Virtanen2016}
{Virtanen} I.,  {Mursula} K.,  2016, \mn@doi [\aap]
  {10.1051/0004-6361/201628096}, \href
  {https://ui.adsabs.harvard.edu/abs/2016A&A...591A..78V} {591, A78}

\bibitem[\protect\citeauthoryear{{Virtanen} \& {Mursula}}{{Virtanen} \&
  {Mursula}}{2019}]{Virtanen2019}
{Virtanen} I.,  {Mursula} K.,  2019, \mn@doi [\aap]
  {10.1051/0004-6361/201935713}, \href
  {https://ui.adsabs.harvard.edu/abs/2019A&A...626A..67V} {626, A67}

\bibitem[\protect\citeauthoryear{{Wang} \& {Panasenco}}{{Wang} \&
  {Panasenco}}{2019}]{Wang2019}
{Wang} Y.~M.,  {Panasenco} O.,  2019, \mn@doi [\apj]
  {10.3847/1538-4357/aaff5e}, \href
  {https://ui.adsabs.harvard.edu/abs/2019ApJ...872..139W} {872, 139}

\bibitem[\protect\citeauthoryear{{Wang} \& {Robbrecht}}{{Wang} \&
  {Robbrecht}}{2011}]{Wang2011}
{Wang} Y.~M.,  {Robbrecht} E.,  2011, \mn@doi [\apj]
  {10.1088/0004-637X/736/2/136}, \href
  {https://ui.adsabs.harvard.edu/abs/2011ApJ...736..136W} {736, 136}

\bibitem[\protect\citeauthoryear{{Wang}, {Sheeley}  \& {Rich}}{{Wang}
  et~al.}{2007}]{Wang2007}
{Wang} Y.~M.,  {Sheeley} N.~R. J.,   {Rich} N.~B.,  2007, \mn@doi [\apj]
  {10.1086/511416}, \href
  {https://ui.adsabs.harvard.edu/abs/2007ApJ...658.1340W} {658, 1340}

\bibitem[\protect\citeauthoryear{{Zhao}, {Hoeksema}  \& {Scherrer}}{{Zhao}
  et~al.}{2005}]{Zhao2005}
{Zhao} X.~P.,  {Hoeksema} J.~T.,   {Scherrer} P.~H.,  2005, \mn@doi [Journal of
  Geophysical Research (Space Physics)] {10.1029/2004JA010723}, \href
  {https://ui.adsabs.harvard.edu/abs/2005JGRA..11010101Z} {110, A10101}

\makeatother
\end{thebibliography}




\appendix
\section{Trajectory of the Parker Solar Probe}

\begin{figure*}
	\centering
	\centerline{\includegraphics[width=0.9\textwidth]{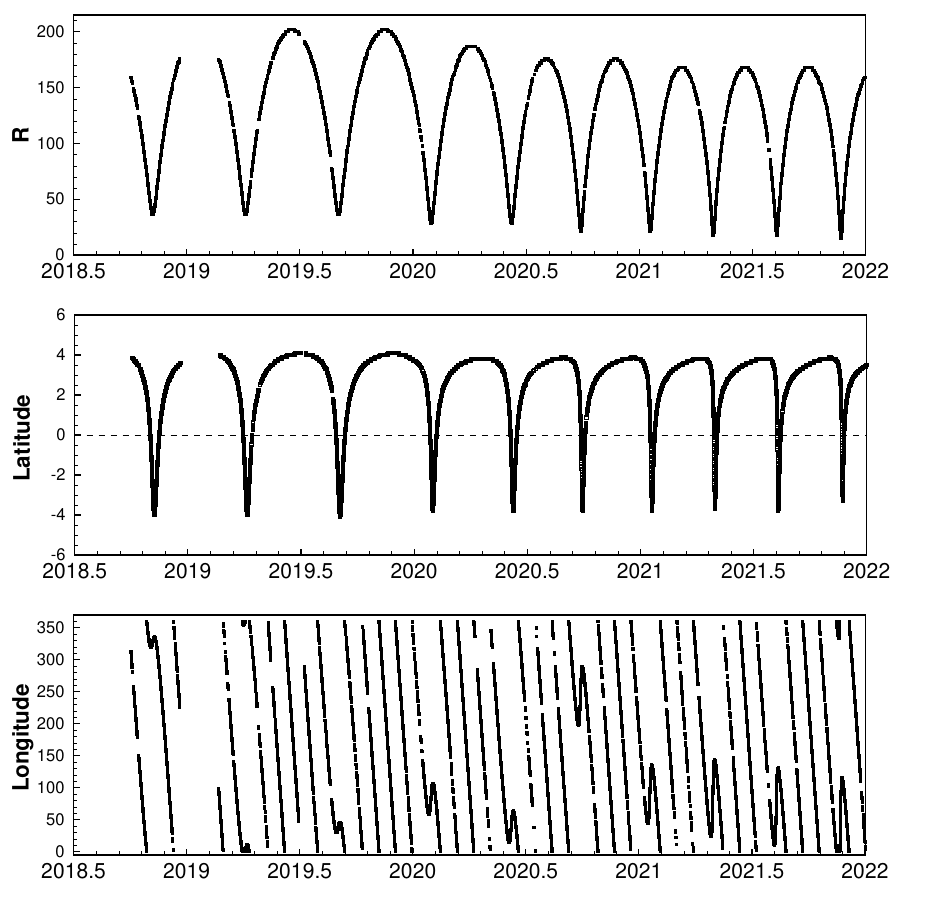}}
	\caption{Heliocentric distance (R), latitude and Carrington longitude of the Parker Solar Probe (PSP). Position where IMF data is not available is shown by data gaps. }
	\label{fig:PSP_orbit}
\end{figure*} 


\bsp	
\label{lastpage}
\end{document}